\documentclass[a4paper]{article}

\usepackage{graphicx}	% Include figure filed
\usepackage{multirow}
\usepackage{xspace}	% Include xspace
\usepackage{amsmath}
\usepackage{amssymb}

\newcommand \sqsn{\mbox{$\sqrt{s_{_{NN}}}$}\xspace}

\newcommand{\UA}{\ensuremath{U_{A}(1)}\xspace}

\newcommand{\etap}{\ensuremath{{\eta^\prime}}\xspace}

\newcommand{\GeV}{\ensuremath{\mathrm{GeV}}\xspace}

\begin{document}
\let\section=\subsection %decrease font size this tricky way

\date{January 11, 2011}

\title{Restoration of a Lost Symmetry \\[10pt] Mass Reduction of the \etap Mesons in $10^{-22}$ Sec}

\author{By T. Cs\"org\H{o}, R. V\'ertesi and J. Sziklai}

\maketitle

\noindent
A particle called \etap meson is found to reduce its mass in less than $10^{-22}$ second in high energy heavy ion collisions at RHIC. Apparently this is the fastest mass reduction ever observed. This indicates the restoration of a lost symmetry of strong interactions in a hot and dense, hadronic matter.

\vspace{10pt}
\noindent
The $\eta$ and the \etap mesons are similar to a pair of identical twins -- their quark content is identical. Under usual circumstances, however, the $\etap(958)$ is nearly twice as heavy as its partner, the $\eta(548)$ meson. High energy heavy ion collisions at Brookhaven National Laboratory's RHIC accelerator generate a hot soup of quark gluon plasma (sQGP) which rehadronizes at temperatures of about 2 Terakelvin, and produces mesons like $\eta$ and \etap
in a fleetingly short time of about $10^{-22}$ sec.

\vspace{10pt}
\noindent
Recently, in a paper that appeared in the Physical Review Letters, we reported on an indirect observation of a significant, at least 200 MeV mass reduction of the \etap mesons~\cite{refone} in the hot and dense, hadronic medium. Such a medium is formed after the quarks and gluons of sQGP are reconfined into mesons and baryons in a process called hadronization. As long as the \etap dwells in such a hot and dense, hadronic medium, the huge mass difference of 410 MeV between the $\eta$ and the \etap mesons disappears within the errors of the analysis, which is based on a combined dataset of the STAR and PHENIX Collaborations. This is just like what might happen in the blink of an eye if the overweight partner of a set of identical twins suddenly lost its extra weight and became the pre-calculated, ideal weight and shape of the slimmer twin.

\vspace{10pt}
\noindent
Such a mass reduction might indicate a restoration of an important symmetry of strong interactions, the so-called \UA symmetry, and the return of a previously lost, "prodigal" Goldston boson, the in-medium modified \etap. Our results indicate that this \UA symmetry is apparently restored in a hot and dense, hadronic matter~\cite{refone}. This symmetry restoration happens at temperatures that are below the temperature range of sQGP formation, in agreement with theoretical predictions based on quark model calculations~\cite{reftwo} and discretized (lattice) quantum chromodynamics~\cite{refthree}.

\newpage

\begin{figure}[h!]
\begin{center}
\includegraphics[width=\linewidth]{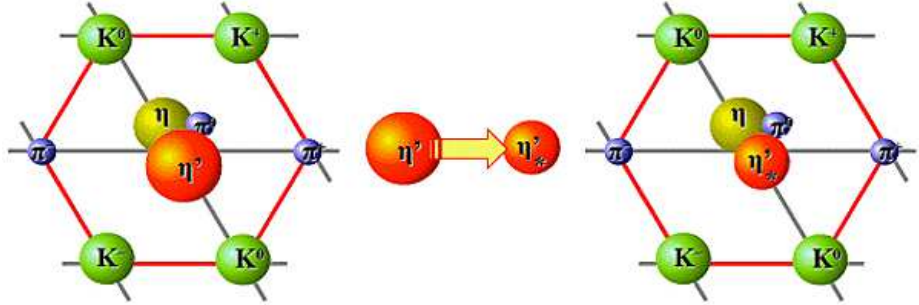}%
\end{center}
\caption{%
The 9 pseudoscalar mesons formed by $u$, $d$ and $s$ quarks are represented as pellets, with sizes proportional to their masses. The left panel indicates a usual situation, corresponding to masses measured in elementary particle induced reactions (Particle Data Group values). The medium plot indicates a mass reduction of the \etap mesons in hot and dense hadronic matter, created in $\sqsn = 200\ \GeV$ Au+Au collisions at the Relativistic Heavy Ion Collider, Brookhaven National Laboratory, USA. The right panel indicates the restoration of a symmetry between the $\eta$ and the $\etap$ mesons in hot and dense hadronic matter.%
}
\end{figure}

\section*{About the Authors}

Tam\'as Cs\"org\H{o}\\
visiting research scholar at Harvard University,\\
Scientific Advisor at MTA KFKI RMKI, Budapest, Hungary\\
\\
R\'obert V\'ertesi\\
PhD student at Debrecen University,\\
affiliated with MTA KFKI RMKI, Budapest, Hungary,\\
supervised by Tam\'as Cs\"org\H{o} and G\'abor D\'avid (BNL)\\
\\
J\'anos Sziklai\\
senior research scholar at MTA KFKI RMKI, Budapest, Hungary

\end{document}